\documentclass{icmart}
\usepackage[dvips]{graphics}
\usepackage{pstricks,pst-plot,pst-node}
\newpsobject{showgrid}{psgrid}{subgriddiv=1,griddots=5,gridlabels=0pt}

\contact[okounkov@math.princeton.edu]{Department of Mathematics, 
Princeton University, Fine Hall, Washington Road, Princeton, New Jersey, 08544, U.S.A.}

\newtheorem{theorem}{Theorem}

\theoremstyle{definition}

\newcommand{\Var}{\operatorname{Var}}

\title[Instanton counting]
{Random partitions and instanton counting}

\author[Andrei Okounkov]{Andrei Okounkov\thanks{The author thanks Packard 
Foundation for partial financial support. }}

\newcommand{\C}{\mathbb{C}}
\newcommand{\Z}{\mathbb{Z}}
\newcommand{\R}{\mathbb{R}}
\newcommand{\Q}{\mathbb{Q}}
\newcommand{\cO}{\mathcal{O}}
\newcommand{\cF}{\mathcal{F}}
\newcommand{\bC}{\mathsf{C}}
\newcommand{\bA}{\mathsf{A}}
\newcommand{\bU}{\mathbf{U}}
\newcommand{\fF}{\mathfrak{F}}
\newcommand{\fM}{\mathfrak{M}}
\newcommand{\bG}{\mathsf{G}}
\newcommand{\bL}{\mathsf{L}}
\newcommand{\fS}{\mathfrak{S}}
\newcommand{\cE}{\mathcal{E}}
\newcommand{\cX}{\mathcal{X}}

\newcommand{\bPs}{\mathbf{\Psi}}

\newcommand{\bM}{\mathsf{M}}
\newcommand{\cM}{\mathcal{M}}
\newcommand{\cS}{\mathcal{S}}

\newcommand{\bP}{\mathbb{P}}

\newcommand{\Mb}{\overline{\mathcal{M}}}

\newcommand{\du}{^{\,\vee}}

\DeclareMathOperator{\Ext}{Ext}

\DeclareMathOperator{\tr}{tr}
\DeclareMathOperator{\Lie}{Lie}

\DeclareMathOperator{\vol}{vol}
\DeclareMathOperator{\diag}{diag}

\newtheorem*{theorem*}{Theorem}

\begin{document}

\begin{abstract}
We summarize the connection between 
random partitions and 
$\mathcal{N}=2$ 
supersymmetric
gauge theories in $4$ dimensions and 
indicate how this relation extends to 
higher dimensions. 
\end{abstract}

\begin{classification}
Primary 81T13; Secondary 14J60.
\end{classification}


\maketitle

\section{Introduction}

\subsection{Random partitions}
\begin{figure}[!hbtp]
  \centering
\scalebox{0.45}{\includegraphics{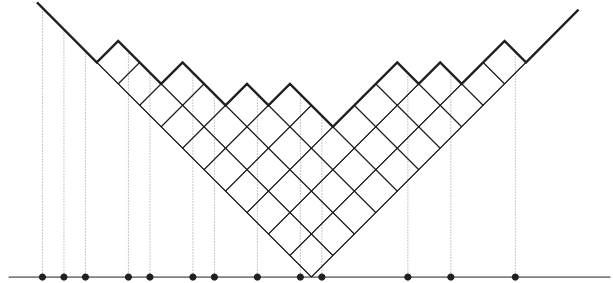}}
\caption{The diagram of $\lambda=(10,8,7,4,4,3,2,2,1,1)$, 
flipped and 
rotated by $135^\circ$. Bullets indicate the 
points of $\fS(\lambda)$. The profile of 
$\lambda$ is plotted in bold.}
  \label{partition}
\end{figure}

A partition of $n$ is a monotone sequence
$$
\lambda = (\lambda_1 \ge \lambda_2 \ge \dots \ge 0)
$$
of nonnegative integers with sum $n$. 
The number $n$ is denoted $|\lambda|$
and called the size of $\lambda$. A geometric object associated
to a partition is its \emph{diagram}; it contains $\lambda_1$
squares in the first row, $\lambda_2$ squares in the second 
row and so on. An example, flipped and rotated by $135^\circ$
can be seen in Figure \ref{partition}. 
Partitions naturally label many basic objects in mathematics
and physics, such as e.g.\ conjugacy classes and representations
of the symmetric group $S(n)$, and very often appear as modest 
summation ranges or indices. A simple but fruitful change of 
perspective, which I wish to stress here, is to treat
sums over partitions probabilistically, that is, treat them as
expectations of some functions of a random partition. 

A survey of the theory of random partitions may be found 
in \cite{uses}. Of the several natural measures on partitions, 
the Plancherel measure
\begin{equation}
  \label{Pla}
   \fM_\textup{Planch}(\lambda) = \frac{(\dim \lambda)^2}{n!} \,, 
\quad |\lambda|=n\,,
\end{equation}
stands out as the one
with deepest properties and widest applications. Here 
$\dim \lambda$ is the dimension of the corresponding representation 
of $S(n)$. This is a probability measure on partitions of $n$.
It can be viewed as a
distinguished discretization of the GUE 
ensemble of the random matrix theory. Namely, a
 measure on partitions can be made a point process on 
a lattice by placing particles in positions 
$$
\fS(\lambda) =\left\{\lambda_i-i+\tfrac12\right\} \subset 
\Z+\tfrac12 \,. 
$$
Figure \ref{partition} illustrates the geometric meaning of 
this transformation. An important theme of recent research 
was to understand how and why  for a Plancherel random partition 
of $n\to\infty$ 
the particles $\fS(\lambda)$ behave like the eigenvalues of a random 
Hermitian matrix. See \cite{BDJ,BOO,J,randpe} and e.g.\ \cite{Jo2,randpar,uses}
for a survey. 

In these notes, we consider a different problem, namely, the behavior 
of \eqref{Pla} 
in a \emph{periodic potential}, that is, additionally weighted
by a multiplicative periodic function of the particles' positions. 
This leads to new phenomena and new applications. As we will see, the 
partition function of $\fM_\textup{Planch}$ in a periodic 
potential is closely related to Nekrasov partition 
function from supersymmetric gauge theory. This relationship will be 
reviewed in detail in Section \ref{sGau} and its 
consequences may be summarized as follows. 

\subsection{Instanton counting}
In 1994, Seiberg and Witten proposed an 
exact description of the low-energy behavior 
of certain supersymmetric gauge 
theories \cite{SW1,SW2}. In spite of the enormous
body of research that this insight 
has generated, only a modest progress was
made towards its gauge-theoretic derivation. 
This changed in 2002, when Nekrasov proposed in 
\cite{N} 
a physically meaningful and mathematically 
rigorous definition of the regularized partition function $Z$
for supersymmetric gauge theories in question. 

Supersymmetry makes the gauge theory partition
function the partition function of a 
\emph{gas of instantons}. Nekrasov's idea was to 
use \emph{equivariant integration} with respect to 
the natural symmetry group in lieu of a long-distance cut-off
for the instanton gas.  He conjectured that as the 
regularization 
parameter $\varepsilon\to 0$
$$
\ln Z \sim - \frac{1}{\varepsilon^2} \cF 
$$
where the free energy $\cF$ expressed by 
the Seiberg-Witten formula in terms of periods
of a certain differential $dS$ on 
a certain algebraic curve $C$. 

This conjecture was proven in 2003 by Nekrasov and 
the author for a list of gauge theories with 
gauge group $U(r)$, namely, pure gauge theory, 
theories with matter fields in fundamental and 
adjoint representations of the gauge group, 
as well as $5$-dimensional theory compactified
on a circle \cite{NO}. Simultaneously, 
independently, and using completely different 
ideas, the formal power series version 
of Nekrasov's conjecture was proven for the pure
$U(r)$-theory by Nakajima and Yoshioka \cite{NY1}. The methods of \cite{NO} 
were applied to classical gauge groups in \cite{ABCD} and
to the 6-dimensional 
gauge theory compactified on a torus in \cite{Iq}. 
Another algebraic approach, which works for pure gauge 
theory with any gauge group,   
was developed by Braverman \cite{Bra} and 
Braverman and Etingof \cite{BraE}.

In these notes, we outline the results of 
\cite{NO} in the simplest, yet fundamental, 
case of pure gauge theory. As should be obvious
from the title, the main idea is to 
treat the gauge theory partition function $Z$ as
the partition function of an ensemble of 
random partitions. 
The $\varepsilon\to 0$ limit turns out to be 
the \emph{thermodynamic limit} in this ensemble. 
What emerges in this limit is 
a nonrandom \emph{limit shape}, 
an example of which may be seen in Figure \ref{limshape}. 
This is a form of the law of large numbers, analogous,
 for example, to Wigner's semicircle law for the spectrum 
of a large random matrix. The limit shape is characterized
as the unique minimizer $\psi_\star$ 
of a certain convex functional $\cS(\psi)$, 
leading to 
$$
\cF = \min \cS \,.
$$ 
We solve the variational problem explicitly and the limit
shape turns out to be an algebraic curve $C$ is disguise. Namely, 
the limit shape is essentially the graph of the function
$$
\Re\int^x_{x_0} dS\,,
$$
where $dS$ is the Seiberg-Witten differential. Thus all ingredients
of the answer appear very naturally in the proof. 

Random matrix theory and philosophy had many successes in 
mathematics and physics. Here we have an example when 
random partitions, while structurally resembling 
random matrices, offer several advantages. First, the 
transformation into a random partition problem is 
geometrically natural and exact. Second, the 
discretization inherent in partitions regularizes
several analytic issues. For further examples along 
these lines the reader may consult \cite{uses}.

\subsection{Higher dimensions}
The translation of the gauge theory problem 
into a random partition problem is explained in 
Section \ref{sGau}. In Section \ref{sPar}, we 
analyze the latter problem, in particular, derive and 
solve the variational problem for the limit 
shape. Section \ref{sHigh} summarizes
parallel results for $3$-dimensional partitions, 
where similar algebraic properties of limit 
shapes are now proven in great generality.  

The surprising fact that free energy $\cF$ is given 
in terms of periods of a hidden algebraic curve $C$
is an example of \emph{mirror symmetry}. 
A general program of interpreting mirror partners 
as limit shapes was initiated in \cite{ORV}. 
Known results about the limit shapes of 
periodically weighted $3$-dimensional 
partitions, together with the conjectural equality 
of Gromov-Witten and Donaldson-Thomas theories
of projective algebraic $3$-folds \cite{mnop}
can be interpreted as a verification of this 
program for toric Calabi-Yau $3$-folds. 
See \cite{EC} for an introduction to these ideas. 

Note that something completely different is 
expected to happen in dimensions $>3$, where
the behavior of both random interfaces and Gromov-Witten 
invariants changes qualitatively.

\section{The gauge theory problem}\label{sGau}

\subsection{Instantons}

We begin by recalling some basic facts, see 
\cite{DK} for an excellent mathematical treatment and 
\cite{DP,many,W} for a physical one. This will serve as motivation 
for the introduction of Nekrasov's partitions function 
in \eqref{defZ} below. 

In gauge theories, 
interactions are transmitted by gauge fields, that is, 
unitary connections on appropriate vector bundles. 
In coordinates, these are matrix-valued functions 
$A_i(x)$ that define covariant derivatives
$$
\nabla_i  = \frac{\partial}{\partial x_i} + A_i(x)\,, \quad 
A_i^* = - A_i \,. 
$$
We consider the most basic case of the 
trivial bundle $\R^4\times \C^r$ over the flat 
Euclidean space-time $\R^4$, where such coordinate 
description is global. 

The natural (Yang-Mills)  
energy functional for gauge fields is 
$L^2$-norm squared $\|F\|^2$ of the curvature
$$
F = \sum \left[\nabla_i,\nabla_j\right] \, dx_i \wedge dx_j \,.
$$
The path integral 
in quantum gauge theory then takes the form 
\begin{equation}
  \label{Fey}
  \int_{\textup{connections}/\mathcal{G}} \mathcal{D}A \,\, 
\exp\left(- \beta \, \|F\|^2\right) \times \dots \,, 
\end{equation}
where dots stand for terms involving other fields of the 
theory  and 
$\mathcal{G}$ is the group of gauge transformations 
$g:\R^4\to U(r)$ acting by 
$$
\nabla \mapsto g \, \nabla \, g^{-1} \,. 
$$
In these notes,
we will restrict ourselves to pure gauge theory, which is already 
quite challenging due to the complicated form of the 
energy. A parallel treatment of certain matter fields
can be found in \cite{NO}.  

Our 
goal is to study \eqref{Fey}
as function of the parameter 
$\beta$ (and boundary conditions at infinity, see below). 
A head-on probabilistic approach to this problem would be 
to make it a theory of many interacting random matrices
through a discretization of space-time. This is a fascinating 
topic about which I have nothing to say. In a different 
direction, when $\beta\gg0$, the minima of $\|F\|^2$
should dominate the integral. In \emph{supersymmetric} gauge 
theory, there is a way to make such approximation exact, 
thereby reducing the path integral to the following 
finite-dimensional integrals. 

Local minima of $\|F\|^2$ 
are classified by a topological invariant $c_2\in \Z$, 
$$
c_2 = \frac1{8\pi^2} \int_{\R^4} \tr F^2\,, 
$$
called \emph{charge}, and satisfy a system of first order PDEs 
\begin{equation}
  \label{ASD}
    F \pm \star F =0 \,, 
\end{equation}
where $\star$ is the Hodge star operator on $2$-forms on $\R^4$. 
With the plus sign, \eqref{ASD} corresponds to $c_2>0$ and 
is called the anti-self-duality equation. Its solutions
are called \emph{instantons}. Minima with $c_2<0$ are 
obtained  by reversing the orientation of $\R^4$. 

The ASD equations \eqref{ASD} are conformally invariant and 
can be transported to a punctured $4$-sphere $S^4 = \R^4 \cup
\{\infty\}$ via stereographic projection. {F}rom 
the removable singularities theorem
of Uhlenbeck it follows that 
 any instanton on $\R^4$ extends, 
after a gauge 
transformation, to an 
instanton on $S^4$. Thus we can talk about the 
value of an instanton at infinity. 

Let $\mathcal{G}_0$ be the group of maps $g:S^4\to U(r)$
such that $g(\infty)=1$.  Modulo $\mathcal{G}_0$, 
instantons on $S^4$ with $c_2=n$ are 
parametrized by a smooth manifold $\cM(r,n)$ 
of real dimension $4 r n$. Naively, 
 one would like the contribution from charge 
$n$ instantons to \eqref{Fey} to be the volume 
of $\cM(r,n)$ in a natural symplectic structure. 
However, $\cM(r,n)$ is noncompact (and its volume is 
infinite) for two following reasons.

Approximately, an element 
of  $\cM(r,n)$ can be imagined as a nonlinear 
superposition of $n$ instantons of charge $1$. 
Some of those may become point-like, i.e.\ their curvature
may concentrate in a $\delta$-function spike, while others
may wander off to infinity. A partial compactification of 
$\cM(r,n)$, constructed by Uhlenbeck, 
which replaces point-like instanton 
by just points of $\R^4$, takes care of the first problem
but not the second.  Nekrasov's idea was to use \emph{equivariant 
integration} to regularize the instanton contributions.

\subsection{Equivariant regularization}
The group 
$$
K = SU(2) \times SU(r) 
$$
acts on $\cM(r,n)$ by rotations of $\R^4=\C^2$ and constant 
gauge transformation, respectively. Our plan is to use 
this action for regularization. 
Let's start with the following simplest example: suppose 
we want to regularize the volume of $\R^2$. A gentle 
way to do it is to introduce a Gaussian well 
\begin{equation}
  \label{Gauss}
    \int_{\R^2} e^{-t \pi (x^2+y^2)} dx \, dy = \frac{1}{t} 
\,, \quad \Re t \ge 0 
\end{equation}
and thus an effective cut-off at the $|t|^{-1/2}$ scale. 
Note that the Hamiltonian flow on $\R^2$ generated by $H=\frac12 (x^2+y^2)$
with respect to the standard symplectic form $\omega=dx\wedge dy$
is rotation about the origin with angular velocity one. 
This makes \eqref{Gauss}
a simplest instance of the 
Atiyah-Bott-Duistermaat-Heckman 
\emph{equivariant localization} formula \cite{AB}. We will use
localization in the following complex form.

Let
$T=\C^*$ act on a complex manifold $X$ with isolated fixed points
$X^T$. Suppose that the action of $U(1)\subset T$ is generated 
by a Hamiltonian $H$ with respect to a symplectic form $\omega$. 
Then 
\begin{equation}
  \label{AB}
    \int_{X} e^{\,\omega - 2 \pi t H} = \sum_{x\in X^T} \frac{e^{-2 \pi t H(x)}}
{\det t|_{T_x X}}\,, 
\end{equation}
where $t$ should be viewed as an element of 
$\Lie(T) \cong \C$, an so it acts in the complex
tangent space $T_x X$ to a fixed point $x\in X$. While \eqref{AB}
is normally stated for compact manifolds $X$, example 
\eqref{Gauss} shows
that with care it can work for noncompact ones, too. 
Scaling both 
$\omega$ and $H$ to zero, we get from \eqref{AB} a formal 
expression
\begin{equation}
  \label{int1}
    \int_{X} 1  \,\overset{\textup{\tiny def}}=\,
 \sum_{x\in X^T} \frac{1}
{\det t|_{T_x X}}\,, 
\end{equation}
which 
does not depends on the symplectic form 
and vanishes if $X$ is compact.

A theorem of Donaldson 
identifies instantons with 
\emph{holomorphic bundles} on $\C^2=\R^4$ and thus 
gives a complex description 
of $\cM(r,n)$. Concretely,  $\cM(r,n)$
is the moduli space of rank $r$ holomorphic 
bundles $\cE\to \C\bP^2$ with given 2nd Chern class $c_2(\cE)=n$ and a 
given trivialization along the line 
$$
L_\infty = \C\bP^2 \setminus \C^2 
$$
at infinity. Note that existence of such trivialization 
implies that $c_1(\cE)=0$. A similar but larger 
moduli space $\Mb(r,n)$
of \emph{torsion-free sheaves}, see e.g.\ \cite{HL,Nb}, is a smooth
partial compactification  of $\cM(r,n)$. 

The complexification of $K$ 
$$
K_\C= SL(2) \times SL(r) 
$$ 
acts on $\Mb(r,n)$ by operating on $\C^2$ and changing 
the trivialization at infinity. 
Equivariant localization 
with respect to a general $t\in\Lie(K)$
\begin{equation}
  \label{gen_t}
     t = (\diag (-i\varepsilon,i\varepsilon), \diag(ia_1,\dots,ia_r))
\end{equation}
combines the two following effects. First, it introduces
a spatial cut-off parameter $\varepsilon$ as in \eqref{Gauss}. 
Second, it introduces dependence on the instanton's 
behavior at infinity through the parameters $a_i$. 
While the first factor in $K$ works to shepherd 
run-away instantons back to the origin, the second 
works to break the gauge invariance at infinity. 
In supersymmetric gauge theories, the parameters 
$a_i$ correspond to the vacuum expectation of the 
Higgs field and thus are responsible for 
 masses of gauge bosons.  In 
short, they are live physical parameters.

\subsection{Nekrasov partition function}
We are now ready to introduce our main object of 
study, the partition function of the
pure ($\mathcal{N}=2$ supersymmetric) $U(r)$ gauge theory: 
\begin{equation}
  \label{defZ}
    Z(\varepsilon;a_1,\dots,a_r;\Lambda) = Z_{\textup{pert}} \,
\sum_{n\ge 0} \Lambda^{2rn} \, \int_{\Mb(r,n)} 1 \,, 
\end{equation}
where the integral is defined by \eqref{int1} 
applied to \eqref{gen_t}, 
$$
\Lambda=\exp(-4\pi^2\beta/r)\,, 
$$
and $Z_{\textup{pert}}$ is a certain perturbative 
factor to be discussed below. The series in \eqref{defZ}
is denoted $Z_{\textup{inst}}$. Because of factorials 
in denominators, see \eqref{Planch}, $Z_{\textup{inst}}$
converges whenever we avoid zero denominators, that is, 
on the complement of 
\begin{equation}
  \label{poles}
    a_i - a_j \equiv 0 \mod \varepsilon \,. 
\end{equation}
In essence, these factorials are there because the 
instantons are unordered. Also note that $Z$ is  
an even function of $\varepsilon$ and 
a symmetric function of the $a_i$'s. 

Since by our regularization rule
$$
\vol \R^4 = \int_{\R^4} 1 = \frac{1}{\varepsilon^2}\,, 
$$ 
we may expect that 
as $\varepsilon\to 0$  
$$
\ln Z(\varepsilon;a;\Lambda) \sim - \frac1{\varepsilon^2} \, \cF
(a;\Lambda)\,, 
$$
where $\cF$ is the \emph{free energy}. At first, the poles
\eqref{poles} of $Z_{\textup{inst}}$, which are getting denser and denser,  
may look like a problem. Indeed, poles of multiplicity 
$O(\varepsilon^{-1})$ may affect the free energy, but 
it is a question of competition with the other terms in $Z_{\textup{inst}}$, 
which the pole-free terms win if $|a_i-a_j| \gg 0$. 
As a result, either by passing to a subsequence of $\varepsilon$, or 
by restricting summation in $Z_{\textup{inst}}$ to the relevant
pole-free terms, we obtain a limit
$$
\cF_\textup{inst} = - \lim \varepsilon^2 Z_\textup{inst}\,,
$$
which 
is analytic and monotone far enough from the 
walls of the Weyl chambers. Recall that Weyl
chambers for $SU(r)$ are the $r!$ 
cones obtained from 
$$
\bC_+ = \left\{ a_1 > a_2 > \dots > a_r, \sum a_i =0  \right\} 
$$
by permuting the coordinates. As $|a_i-a_j|$ get small, 
poles do complicate the asymptotics. This is the origin of 
cuts in the analytic function $\cF(a)$, $a_i\in \C$. 

Nekrasov conjectured in \cite{N} that the free energy 
$\cF$ is the 
\emph{Seiberg-Witten prepotential}, first obtained
in \cite{SW1,SW2} through entirely different considerations. 
It is defined in terms of a certain 
family of algebraic curves.  

\subsection{Seiberg-Witten geometry}\label{sSW} 
In the affine space of complex 
polynomials of the form $P(z) = z^r + O(z^{r-2})$
consider the open set $\bU$ of 
polynomials 
such that 
\begin{equation}
  \label{PzL}
    P(z) = \pm 2 \Lambda^r 
\end{equation}
has $2r$ distinct roots. Over $\bU$, we have a $g$-dimensional
family of complex algebraic 
curves $C$ of genus $g=r-1$ defined by 
\begin{equation}
  \label{SW}
    \Lambda^r \left(w + \frac1{w}\right) = P(z)\,, \quad P\in \bU\,. 
\end{equation}
The curve \eqref{SW} is compactified by 
adding two points  $\partial C =\{w=0,\infty\}$.
\begin{figure}[!hbtp]
  \centering
\scalebox{0.55}{\includegraphics{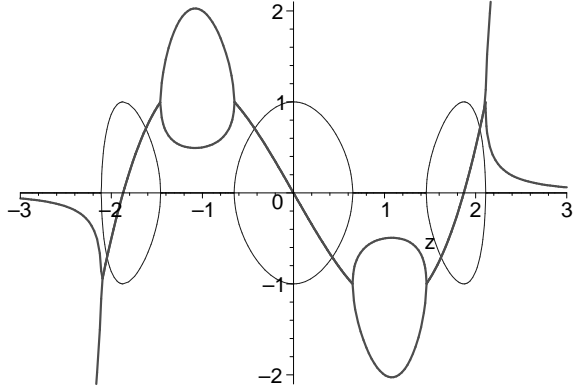}}
\caption{$\Re w$ (bold) and $\Im w$ for 
$w+1/w=z^3-3.5 \, z$ and $z\in\R$}
  \label{ovals}
\end{figure}

Let $\bM\subset\bU$ be the set of $P(z)$ for 
which all roots of $\eqref{PzL}$ are real. 
The corresponding curves $C$ are called \emph{maximal} 
and play a special role, see e.g.\ \cite{SY}. They 
arise, for example, as spectral curves
of a periodic Toda chain \cite{Toda}. 
A maximal curve $C$ has $r$ real ovals, as 
illustrated in Figure \ref{ovals}. 
Note that for $z\in\R$, $w$ is either real 
or lies on the unit circle $|w|=1$.

The intervals 
$P^{-1} ([-2\Lambda^r,2\Lambda^r]) \subset \R$ 
on which $|w|=1$ are 
called \emph{bands}. The intervals between the 
bands are called \emph{gaps}. 
The smaller (in absolute value) root 
$w$ of the equation \eqref{SW} can be 
unambiguously defined for $z\in \C \setminus 
\{\textup{bands}\}$. On the corresponding 
sheet of the Riemann surface of $w$, we define
cycles
\begin{equation}
  \label{albe}
  \alpha_i \in H_1(C-\partial C)\,, 
\quad  
\beta_i \in H_1(C,\partial C)\,, 
\quad i=1,\dots,r
\end{equation}
as illustrated
in Figure \ref{cycles}, where dotted line means that
$\beta_i$ continues on the other sheet.  
Note 
that $\alpha_i \cap \beta_j = \delta_{ij}$ 
and that 
\begin{equation}
  \label{alba}
    \overline{\alpha}_i = - \alpha_i\,, \quad   
\overline{\beta}_i = \beta_i \,, 
\end{equation}
where bar stands for complex conjugation. 
The ovals in Figure \ref{ovals} represent 
the cycles $\alpha_i$ and $\beta_i - \beta_{i+1}$. 
\psset{unit=0.25 cm}
\begin{figure}[!htbp]
  \centering
\begin{pspicture}(0,0)(36,14)
\rput[lb](0,0){\scalebox{0.75}{\includegraphics{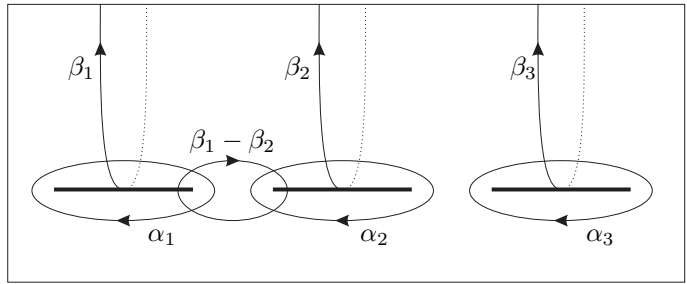}}}
\rput[lb](7.5,2){$\alpha_1$}
\rput[lb](18.8,2){$\alpha_2$}
\rput[lb](30.8,2){$\alpha_3$}
\rput[rt](4.7,12){$\beta_1$}
\rput[rt](16.2,12){$\beta_2$}
\rput[rt](28.2,12){$\beta_3$}
\rput[c](12,7.5){$\beta_1-\beta_2$}
\end{pspicture}
\caption{Cycles $\beta_i$ go from $w=\infty$ to $w=0$. Bold segments
indicate bands.} 
  \label{cycles}
\end{figure}

The Seiberg-Witten differential 
$$
dS = \frac{1}{2\pi i} \, z \, \frac{dw}{w}  = 
\pm \frac{r}{2\pi i} \left(1 + O\left(z^{-2}\right)
\right) \, dz 
$$
is holomorphic except for a second order pole 
(without residue) at $\partial C$. Its derivatives
with respect to $P\in \bU$ are, therefore, holomorphic 
differentials on $C$. In fact, this gives
$$
T_P \bU \cong \textup{holomorphic diff.\ on $C$} \,.
$$
Nondegeneracy of periods implies 
the functions 
\begin{equation}
  \label{ai}
 a_i 
\,\overset{\textup{\tiny def}}=\, 
\int_{\alpha_i} dS \,, \quad \sum a_i = 0 \,, 
\end{equation}
which are real on $\bM$ by \eqref{alba},  are
local coordinates on $\bU$, as are 
\begin{equation}
  \label{adu}
  a\du_{i} - a\du_{i+1}  
\,\overset{\textup{\tiny def}}=\, 
2\pi i \int_{\beta_{i} - \beta_{i+1}} dS \,, \quad \sum a\du_i =0\,. 
\end{equation}
Further, there exists
a function $\cF(a;\Lambda)$, which is real and convex on 
$\bM$, 
such that 
\begin{equation}\label{dper}
\left(
\frac{\partial}{\partial a_i} - 
\frac{\partial}{\partial a_{i+1}} \right)
\cF = - \left(a\du_{i} - a\du_{i+1}\right) \,.
\end{equation}
Indeed, the Hessian of $\cF$ equals $(-2\pi i)$ times the 
period matrix of $C$, hence symmetric (and 
positive definite on $\bM$). The function 
$\cF$ is called the \emph{Seiberg-Witten 
prepotential}. Note that $\cF$ is 
multivalued on $\bU$ and, in fact, its monodromy 
played a key role in the argument of Seiberg and 
Witten. By contrast, $\bM$ is simply-connected, indeed 
$$
a\du: \bM \to \bC_+
$$
is a diffeomorphism, 
see e.g.\ \cite{KO1} for a 
more general result. Note that the periods \eqref{adu}
are the areas enclosed by the images of real ovals of $C$ under
$(z,w)\mapsto (z,\ln|w|)$. 
A similar geometric interpretation 
of the $a_i$'s will be given in \eqref{inter}
below. In particular, the range 
$$
\bA = a(\bM) 
$$
of the 
coordinates \eqref{ai} is a proper subset of $\bC_- = - \bC_+$. 
At infinity of $\bM$, 
we have
$$
a_i \sim \{\textup{roots of $P$}\}\,, \quad  a_1 \ll a_2 \ll
\dots \ll a_r \,.
$$

\subsection{Main result}

We have now defined all necessary ingredients 
to confirm Nekrasov's 
conjecture in the following strong form: 

\begin{theorem}[\cite{NO}] For $a\in \bA$, 
\begin{equation}
    \label{MT}
- \lim_{\varepsilon\to 0} \varepsilon^2 \ln Z(\varepsilon;a;\Lambda) = \cF(a;\Lambda)\,,
\end{equation}
where $\cF$ is the Seiberg-Witten prepotential \eqref{dper}. 
\end{theorem}

At the boundary of $\bA$, free energy has a singularity of the 
form 
$$
\cF = -\left(a\du_i - a\du_j\right)^2 \ln \left(a\du_i - a\du_j\right) + \dots 
$$
where dots denote analytic terms. This singularity is one of the 
main physical features of the Seiberg-Witten theory. 

In broad strokes, the logic of the proof was explained in 
the Introduction. We now proceed with the details.

\section{The random partition problem}\label{sPar}

\subsection{Fixed points contributions}

A rank 1 torsion-free sheaf on $\C^2$ is a fancy name to 
call an ideal $I$ of $\C[x,y]$. Any partition $\lambda$ 
defines one by 
$$
I_\lambda = (x^{\lambda_1},x^{\lambda_2}\, y, x^{\lambda_3}\, y^2,
\dots) \subset \C[x,y] \,.
$$
It is easy to see that all torus-fixed points of $\Mb(r,n)$
have the form 
\begin{equation}
  \label{fF}
    \fF = \bigoplus_{k=1}^r I_{\lambda^{(k)}}\,, \quad 
\sum |\lambda^{(k)}| = n \,,
\end{equation}
where $\lambda^{(k)}$ is an $r$-tuple of partitions. Our goal 
now is to compute the character of the torus action in 
the tangent space to the fixed point \eqref{fF} and thus the 
contribution of 
$\fF$ to the sum in \eqref{int1}. 

By construction 
of $\Mb(r,n)$, its tangent space at $\fF$ 
equals $\Ext^1_{\bP^2}(\fF,\fF(-L_\infty))$. 
{F}rom the vanishing of the other $\Ext$-groups we conclude 
\begin{equation}
  \label{tr_Ext}
     \tr e^t|_{\Ext^1_{\bP^2}(\fF,\fF(-L_\infty))} =
\cX_{\cO^{\oplus r}}(t) - \cX_{\,\fF}(t)\,,
\end{equation}
where $\cX_{\,\fF}(t)$ is the character 
$$
\cX_{\,\fF}(t) = \tr e^t|_{\chi_{\C^2}(\fF,\fF)} 
$$
of the infinite-dimensional virtual representation 
$$
\chi_{\C^2}(\fF,\fF) =  \Ext^0_{\C^2}(\fF,\fF) - 
 \Ext^1_{\C^2}(\fF,\fF) +  \Ext^2_{\C^2}(\fF,\fF) \,.
$$
Any graded free resolution of $\fF$ gives
$$
\mathcal{X}_{\,\fF}(t) =   
|\bG_{\,\fF}(t) |^2\,, \quad t\in \Lie(K)\,,
$$
where 
$\bG_{\,\fF}(t)$ is, up to a factor, the character 
of $\fF$ itself 
\begin{align}
    \bG_{\lambda^{(1)},\dots,\lambda^{(r)}}(t) &= 
(e^{-i\varepsilon\!/2}-e^{i\varepsilon\!/2}) 
\tr e^t|_{\fF}\notag \\ &= 
\sum_{k=1}^r e^{ia_k} \sum_{j=1}^\infty 
\exp\!\left(i \varepsilon (\lambda^{(k)}_j-j+\tfrac12)\right) \,.
\label{bF}
\end{align}
It is also a natural generating function of the 
$r$-tuple $\lambda^{(k)}$. 

Note that the weight of any $\fF$ is real and 
positive, being a product of purely imaginary 
numbers in conjugate pairs.

\subsection{Perturbative factor}
In the spirit of the original uncompactified 
gauge theory problem on $\R^4$, we would like 
to drop the first term in \eqref{tr_Ext} and declare its 
contribution canceled by $Z_\textup{pert}$. In 
view of \eqref{bF}, this requires a regularization 
of the following product 
$$
Z_\textup{pert} \, \textup{``$=$''} \,  
\prod_{k,k'=1}^r \, \prod_{j,j'=1}^\infty i(a_k-a_{k'}+\varepsilon(j-j')) \,.
$$
A natural regularization is provided by Barnes' double
$\Gamma$-function \eqref{douG}, see e.g. \cite{Ruij}.
For $c_1,c_2\in\R$ and $\Re w \gg 0$, define
$$
\zeta_2(s; w\,|\, c_1,c_2) = \frac{1}{\Gamma(s)} \int_0^\infty 
\frac{dt}{t} \, t^s \, \frac{e^{-wt}}{\prod (1-e^{-c_i t})} \,.
$$
This has a meromorphic continuation in $s$ with poles at $s=1,2$. Define 
\begin{equation}
  \label{douG}
\Gamma_2(w \,|\,c_1,c_2) = \exp \left. \frac{d}{ds} \,\zeta(s; w \,|\,
 c_1,c_2)
\right|_{s=0} \,. 
\end{equation}
Through the difference equation 
\begin{equation}
  \label{GaDE}
w \, \Gamma_2(w) \, 
\Gamma_2(w+c_1+c_2) = \Gamma_2(w + c_1) \, \Gamma_2(w + c_2)
\end{equation}
it extends to a meromorphic function of $w$. We define 
\begin{align}
Z_\textup{pert}  =
\prod_{k,k'} 
\Gamma_2\left(\frac{i(a_k-a_{k'})}
{\Lambda}\left|\frac{i\varepsilon}{\Lambda},\frac{-i\varepsilon}{\Lambda}
\right.\right)^{-1}\,. 
\label{Z_pert}
\end{align}
where $\Gamma_2$ is 
analytically continued to imaginary arguments using 
$$
\Gamma_2\left(M w \left| \, M c, - M c\right.\right) = M^{\tfrac{w^2}{2c^2} - 
\tfrac 1{12}} \,\,  \Gamma_2\left(w \left|\, c, - c\right.\right)\,, 
\quad M\notin (-\infty,0] \,. 
$$
The scaling by $\Lambda$ is 
introduced in \eqref{Z_pert} to make \eqref{defZ}
homogeneous of degree $0$ in $a$, $\varepsilon$, and $\Lambda$. 
Note also 
$$
\Gamma_2(0|\,1 , - 1)=e^{-\zeta'(-1)}\,.
$$
Our renormalization rule \eqref{Z_pert}
fits nicely with the following transformation of the
partition function $Z$. 

\subsection{Dual partition function} \label{sZdu} 
For $r=1$, the weight of $I_\lambda$ in \eqref{defZ}
equals 
\begin{equation}
  \label{Planch}
\Lambda^{2n} \,\, {\det}^{-1} t\Big|_{T_{I_\lambda} \Mb(1,n)} = 
\frac{1}{n!} 
\left(\frac{\Lambda^2}{\varepsilon^{2}}
\right)^{\!n}  \fM_\textup{Planch}(\lambda)\,,
\end{equation}
where $\fM_\textup{Planch}$ is the Plancherel measure 
\eqref{Pla} and the prefactor is 
the Poisson weight with parameter $\Lambda^2/\varepsilon^2$. 
For $r>1$, we 
will transform $Z$ into the partition function 
\eqref{per_Pl}
of the Plancherel measure in a \emph{periodic potential}
with period $r$.

Let a function 
$\xi:\Z+\frac12 \to \R$ be periodic with period $r$ and 
mean $0$. The energy $\Xi(\lambda)$ of the configuration 
$\fS(\lambda)$ in the potential $\xi$ is defined by Abel's rule 
$$
\Xi(\lambda) = \sum_{x \in \fS(\lambda)} \, \xi(x) 
\,\overset{\textup{\tiny def}}=\, \lim_{z\to +0}  
\sum_{x \in \fS(\lambda)} \, \xi(x) \, e^{zx} \,. 
$$
Grouping the points of $\fS(\lambda)$ modulo 
$r$ uniquely determines an $r$-tuple of partitions $\lambda^{(k)}$, 
known as $r$-quotients of $\lambda$, and shifts $s_k\in \Q$ 
such that 
\begin{equation}
  \label{quo}
    \fS(\lambda) = \bigsqcup_{k=1}^{r} r \left(\fS
\left(\lambda^{(k)}\right) + s_k \right)
\end{equation}
and 
$$
r s \equiv \rho \mod r\,\Z^r_0 \,, \quad \rho = \left(\tfrac{r-1}{2}, 
\dots, \tfrac{1-r}{2}\right) \,, 
$$
where $\Z^r_0$ denotes vectors with zero sum. 
It follows from \eqref{quo} that 
\begin{equation}
  \label{Ftrans}
    \bG_{\lambda} \left({\varepsilon}/r\right) = 
\bG_{\lambda^{(1)},\dots,\lambda^{(r)}}(\varepsilon;\varepsilon s) \,. 
\end{equation}
Letting $\varepsilon\to 2\pi i k$, $k=1,\dots,r-1$, in \eqref{Ftrans} 
gives 
\begin{equation}
  \label{Xis}
    \Xi(\lambda) = 
(s,\xi) = \sum s_i \, \xi_i\,, \quad \xi_i= \xi\left(\tfrac12-i\right) \,, 
\end{equation}
while the $\varepsilon\to 0$ limit in \eqref{Ftrans} yields 
$$
|\lambda| = r \left(\sum \left|\lambda^{(k)}\right| 
+ \sum \frac{s_k^2}2 \right) 
+ \frac{1-r^2}{24} \,.
$$
Using these formulas and the difference equation \eqref{GaDE},
 we compute 
\begin{align}
Z\du(\varepsilon;\xi_1,\dots,\xi_r;\Lambda)
\,\overset{\textup{\tiny def}}=\, &\sum_{a\in \varepsilon(\rho+r\Z^r_0)} 
\exp\left(\frac{(\xi,a)}{r\varepsilon^2}
\right) \, Z(r\varepsilon; a; \Lambda) 
\label{defZD}\\
= 
e^{\zeta'(-1)+\frac{\pi i}{24}}
&\sum_\lambda \left|\frac{\Lambda}{\varepsilon}\right|^{2|\lambda|
-\frac1{12}} 
\left(\frac{\dim \lambda}{|\lambda|!}\right)^2  
\exp\left(\frac{\Xi(\lambda)}{\varepsilon}
\right)\,. 
\label{per_Pl}
\end{align}
We call \eqref{defZD} the \emph{dual partition function}. By 
\eqref{per_Pl}, it equals the partition function of a
periodically weighted Plancherel 
measure on partitions.

While it will play no role in what follows, it may 
be mentioned here that $Z\du$ is a very interesting 
object to study not asymptotically but exactly. 
For example, Toda equation for $\ln \Z\du$ 
may be found in Section 5 of \cite{NO}.

\subsection{Dual free energy}
Define the dual free energy by 
\begin{equation}
  \label{Fd}
  \cF\du(\xi;\Lambda) = - \lim_{\varepsilon\to 0} \varepsilon^2 \ln Z\du  \,.
\end{equation}
Since \eqref{defZD} is a Riemann sum for 
Laplace transform, we may expect that 
\begin{equation}
  \label{FLeg}
  \cF\du(\xi;\Lambda) = 
\min_{a\in \R_0^r} \frac1{r^2} \, \cF(a;\Lambda) - \frac1r \, (\xi,a)
\end{equation}
that is, up to normalization, $\cF\du$ is 
the Legendre transform of $\cF$. 
This is because the asymptotics of Laplace transform is 
determined by one point --- the maximum. Our plan is 
apply to same logic to the infinite-dimensional
sum \eqref{per_Pl}, namely, 
to show that its $\varepsilon\to 0$ asymptotics 
is determined by a single term, the \emph{limit shape}.

The law of large numbers, a basic principle of 
probability, implies that on a large scale most random system 
are deterministic: solids have definite shape, 
fluids obey the laws of hydrodynamics, etc. Only 
magnification reveals the full randomness of nature.
 
In the case at hand, the weight of a partition
$\lambda$ in \eqref{per_Pl}, normalized by the 
whole sum, defines a probability measure on the 
set of partitions. This measure depends on a 
parameter $\varepsilon$ and as $\varepsilon\to 0$ it clearly favors
partitions of larger and larger size. In fact, the 
expected size of $\lambda$ grows as $\varepsilon^{-2}$. 
We thus expect the 
diagram of $\lambda$, scaled by $\varepsilon$
in both directions, to satisfy a law of large numbers, 
namely, to have a nonrandom limit shape.  
By definition, this limit shape will dominate the leading $\varepsilon
\to 0$ asymptotics 
of $Z\du$. In absence of the 
periodic potential $\Xi$, such analysis is a classical  
result of  Logan-Shepp and Vershik-Kerov \cite{LS,VK1,VK2}.

Note that the maximum in \eqref{FLeg} is over all of 
$a$, including the problematic 
region where $|a_i-a_j|$ get small. However, 
this region does not contribute to $\cF\du$ as the convexity 
of free energy is lost there. We will see this 
reflected in the following properties of $\cF\du$:
it is strictly concave, analytic in the 
interior of the Weyl chambers, and singular
along the chambers' walls. 

\subsection{Variational problem for the limit shape}

The \emph{profile} of a partition $\lambda$ 
is, by definition, the piecewise linear function 
plotted in bold in Figure \ref{partition}. Let $\psi_\lambda$ 
be the profile of $\lambda$ scaled by $\varepsilon$ in 
both directions. The map 
$\lambda\mapsto\psi_\lambda$ 
 embeds partitions into the convex set 
$\bPs$ of functions $\psi$ on $\R$ with 
Lipschitz constant $1$ and 
$$
|\psi| = \int |\psi(x)-|x|| \, dx  < \infty\,. 
$$ 
The Lipschitz condition implies 
$$
\|\psi_1 - \psi_2\|_{C} \le \|\psi_1 - \psi_2\|^{1/2}_{L^1}\,,
\quad \psi_1,\psi_2 \in \bPs\,,
$$
and so $\bPs$ is complete and separable in the $L^1$-metric. 
Some function of a partition have a natural continuous 
extension to $\bPs$, for example
$$
\bG_\lambda(\varepsilon) = \frac1{e^{i\varepsilon/2}-e^{-i\varepsilon/2}} 
\left(1-\frac 12 
\int e^{ix} \, (\psi_\lambda(x)-|x|)\, dx \right)\,, 
$$
while others, specifically the ones appearing in 
\eqref{per_Pl}, do not. An adequate language for 
dealing with this is the following. 

Let $f(\lambda)\ge 0$ be a function on partitions
depending on the parameter $\varepsilon$. We say that it
satisfies a \emph{large deviation} principle with 
action (rate) functional $\cS_f(\psi)$ if for any set 
$A\subset \bPs$ 
\begin{equation}
  \label{LD}
- \lim \varepsilon^2 \, \ln \sum_{\psi_\lambda\in A} f(\lambda)  
\subset \left[
\inf_{\overline{A}} \cS_f, 
\inf_{A^\circ} \cS_f
\right]
\subset \R \cup \{+\infty\}\,,
\end{equation}
where $\lim$ denotes all limit points, $A^\circ$ and 
$\overline{A}$ stand for the interior and closure of $A$, 
respectively. 

For the Plancherel weight \eqref{Planch}, 
Logan-Shepp and Vershik-Kerov proved 
a large 
deviation principle with action 
\begin{equation}
  \label{Spl}
    \cS_{\textup{pl}}(\psi) =  \frac12 
\int_{x <y} (1+\psi'(x))(1-\psi'(y)) \, \ln \frac{|x-y|}{\Lambda} 
\, 
dx \, dy \,.
\end{equation}
Note that in this case  the sum in \eqref{LD} may be 
replaced by maximum because the number of partitions of 
$n$ grows subexponentially 
in $n$. In other words, there is no entropic contribution in \eqref{Spl}. 

\begin{figure}[!hbtp]
  \centering
\scalebox{0.4}{\includegraphics{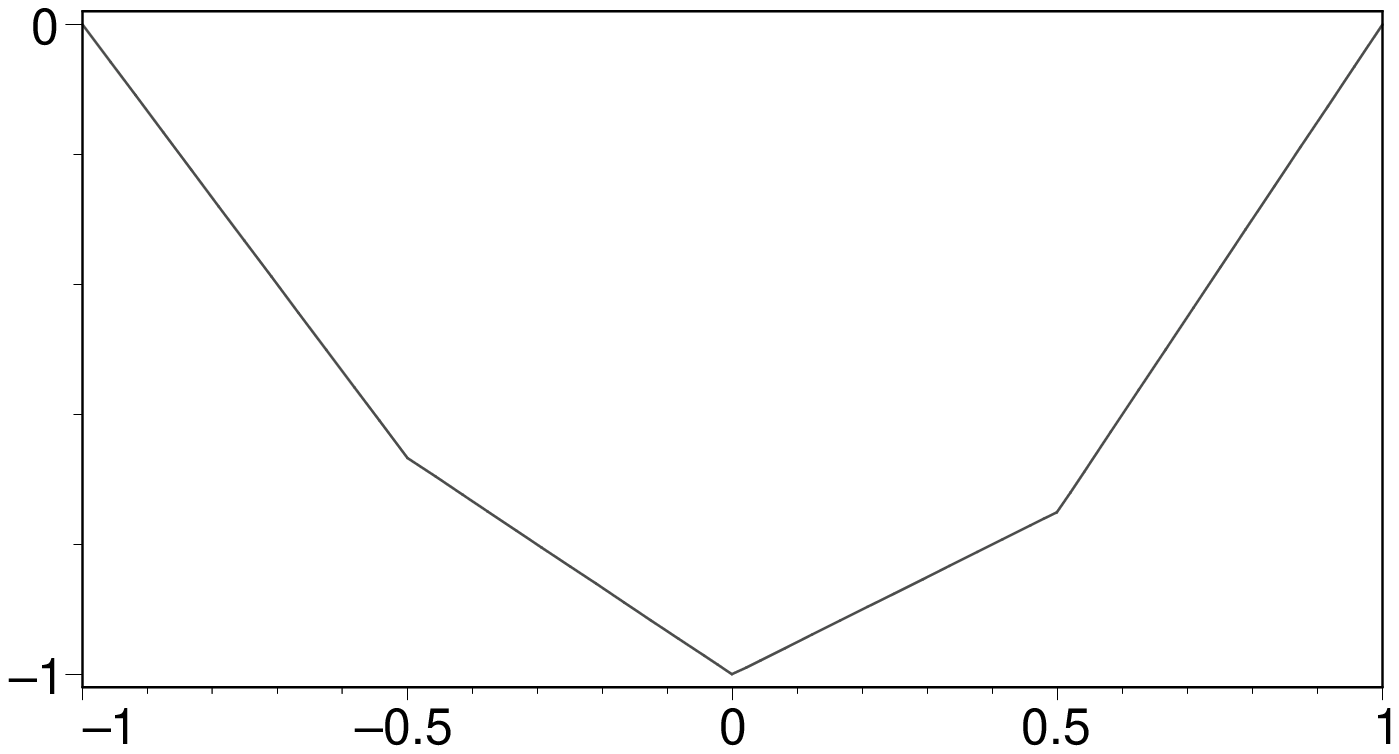}}
\caption{The surface tension $\sigma$ for $r=4$ and $\xi=
\left\{-\frac43,-\frac23,\frac12,\frac32\right\}$}
  \label{surftens}
\end{figure}

The periodic potential $\Xi(\lambda)$ produces a
\emph{surface tension} addition to the total action $\cS$ 
$$
\cS = \cS_{\textup{pl}} +  \cS_{\textup{surf}}\,, \quad 
\cS_{\textup{surf}}(\psi)= 
\frac12 \int \sigma(\psi') \, dt\,,
$$
where $\sigma$ is a convex piecewise-linear function of the 
kind plotted in Figure \ref{surftens}. It is linear on 
segments of length $2/r$ with slopes $\{\xi_i\}$, 
in increasing order. The form of $\cS_{\textup{surf}}$ is 
easy to deduce directly; it can also be seen 
as e.g.\ the most degenerate case of the 
surface tension formula from Section \ref{sPw}. 
The singularities in the 
surface tension $\sigma$ are responsible for 
\emph{facets}, that is, linear pieces, in the minimizer, 
see Figure \ref{limshape}. The slopes of these facets
are precisely the points where $\sigma'$ is 
discontinuous\footnote{
There are many advantages 
in viewing random partitions as $2$-dimensional slices
of random $3$-dimensional objects discussed in Section 
\ref{sHigh}.  From the probability 
viewpoint, this links random partitions with rather 
realistic models of crystalline surfaces with local 
interaction, enriching both techniques and intuition. 
In particular, coexistence of facets and curved regions
in our limit shapes is the same phenomenon as observed
in natural crystals. {F}rom the gauge theory viewpoint, 
it is also very natural, especially in the context of 
$5$-dimensional theory on $\R^4\times S^1$, 
which corresponds to the $K$-theory 
of the instanton moduli spaces. 
}. Note that $\sigma$ and hence $\cS$ is a symmetric 
function of the $\xi_i$'s. 

The functional $\cS$ is strictly 
convex and its sublevel sets 
$\cS^{-1}((-\infty,c])$ are compact, which can be seen 
by rewriting $\cS_{\textup{pl}}$ in terms of 
the Sobolev $H^{1/2}$ norm, see \cite{LS,VK2}. Therefore 
it has a unique minimum $\psi_\star$ --- the limit shape. 
The large deviation principle 
and the definition of the dual free energy and \eqref{Fd} 
together imply 
\begin{equation}
 \cF\du(\xi;\Lambda) = \cS(\psi_\star) \,.
 \end{equation}
Our business, therefore, is to find this minimizer $\psi_\star$.

\subsection{The minimizer}

By convexity, a local minimum of $\cS$ is automatically a
global one. Since $\sigma$ has one-sided derivatives, 
a local minimum can be characterized by nonnegativity
of all directional derivatives. This leads to the 
following \emph{complementary
slackness} conditions for the convolution of 
$\psi_\star'' (x)$ with the kernel 
$$
\bL(x) = x \ln \frac{|x|}{\Lambda} - x  = \int_0^x \ln \left|
\frac{y}{\Lambda}\right| \, dy\,. 
$$
There exists a constant $c_0$, which is the Lagrange multiplier from 
the constraint $\int \delta \psi' = 0$, such that 
\begin{alignat}{2}
 \notag
  \bL * \psi_\star'' (x) + c_0 &= 
\xi_i\,, \qquad &&\psi_\star'(x)\in \left(-1+\tfrac{2i-2}{r},-1+\tfrac{2i}{r}
\right)
\,,\\ 
 \label{conv}
  \bL * \psi_\star'' (x) + c_0 &\in [\xi_i,\xi_{i+1}]\,, 
\qquad &&\psi_\star'(x) = -1+\tfrac{2i}{r}\,,
\end{alignat}
where to simplify notation we assumed that
$$
\xi \in \bC_-\,, \quad \xi_0 = -\infty\,, \quad \xi_{r+1} = +\infty \,.
$$
Recall that $\bC_-$ denotes the negative Weyl chamber. 

The function $\psi_\star''$ will turn out to be 
nonnegative and supported on a 
union of $r$ intervals, which are
precisely the \emph{bands} of Section \ref{sSW}.
The gaps will produce the facets in the limit shape. 

\begin{figure}[!hbtp]
  \centering
\begin{pspicture}(0,0)(27,14)
\rput[lb](-3,0){
\scalebox{0.55}{\includegraphics{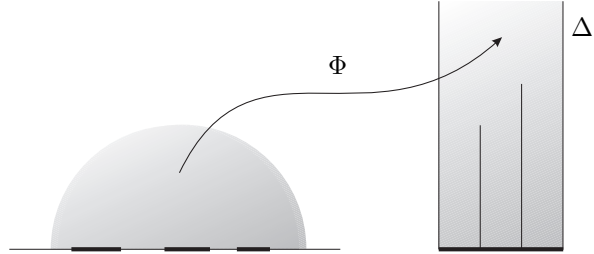}}}
\rput[c](15,10){$\Phi$}
\rput[c](28,12){$\Delta$}
\end{pspicture}
\caption{Conformal map defined by a maximal curve}
  \label{confmap}
\end{figure}

It is elementary to see that for a maximal curve \eqref{SW}
the map 
\begin{equation}
  \label{normPhi}
  \Phi(z) = 1 + \frac{2}{\pi i r} \, \ln w =  
1 + \frac{2}{\pi i} \ln \frac{\Lambda}{z} + 
O(z^{-1})\,, \quad z\to\infty 
\end{equation}
where $w$ is the smaller root of \eqref{SW}, defines a 
conformal map of the upper half-plane to a slit 
half-strip 
$$
\Delta \subset \{z\left| \Im z > 0\,, |\Re z| < 1\right.\}
$$
as in Figure \ref{confmap}. The slits in $\Delta$ go along 
$$
\Re z = -1 + 2i/r\,, \quad i=1,\dots,r-1\,, 
$$
and their lengths are, essentially, the 
critical values of the polynomial $P(z)$. The bands and gaps 
are preimages of the horizontal and vertical segments 
of $\partial \Delta$, respectively. 

\begin{figure}[!htbp]
  \centering
\scalebox{0.45}{\includegraphics{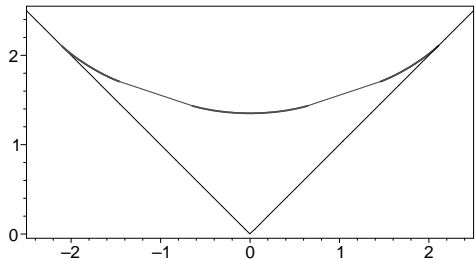}}
\caption{Limit shape corresponding to the curve from 
Fig.\ \ref{ovals}. Thin segments are facets.}
  \label{limshape}
\end{figure}

We claim that 
\begin{equation}
  \label{psis}
    \psi'_\star= \Re \Phi \Big|_{\R} \,, 
\end{equation}
where the polynomial $P(z)$ is determined by the 
relation \eqref{xia} below. 
The equations \eqref{conv} are verified for \eqref{psis}
as follows. Since $\Phi'(z) = O(z^{-1})$, $z\to \infty$,
we have the Hilbert transform relation 
$$
\textup{P.V.} \, \frac{1}{x} * \Re \Phi'\Big|_\R  = 
\pi \Im \Phi'\Big|_\R \,.
$$
Integrating it once and using \eqref{normPhi} to fix
the integration constant, we get 
$$
\left(\bL * \Re \Phi'\right)' = \pi \Im \Phi \,.
$$
Therefore, the function $\bL * \,\Re \Phi'$ is constant on the 
bands and strictly increasing on the gaps, hence \eqref{psis}
satisfies \eqref{conv} with 
\begin{equation}
  \label{xiper}
\xi_{i+1}-\xi_i = \pi \int_{\textup{$i$th gap}} \Im \Phi(x) \, dx
\end{equation}
Integrating \eqref{xiper} by parts and using definitions 
from Section \ref{sSW} gives 
\begin{equation}
  \label{xia}
    \xi = - \frac{a\du}{r}  \,, 
\end{equation}
thus every limit shape $\psi_\star$ comes from a maximal 
curve. For example, the limit shape corresponding to 
the curve from Figure \ref{ovals} is plotted in Figure \ref{limshape}. 
Note also that for $C\in\bM$, we have 
\begin{equation}
  \label{inter}
a_i = \frac{r}{2} \left(I_{i-1} - I_i\right) \,, 
\end{equation}
where $I_i$ is the intercept of the $i$th facet of
the limit shape. In particular, $\bA \subset \bC_-$. 

For given $\xi\in \bC_-$, consider the distribution 
of the $r$-quotients $\lambda^{(i)}$ of the partition 
$\lambda$, as defined in Section 
\ref{sZdu}.  For the  shifts $s_k$ in \eqref{quo} 
we have using \eqref{Xis} 
$$
\varepsilon s \to - \frac{\partial \cF\du}{\partial \xi}\,, 
\quad \varepsilon\to 0 \,,
$$
in probability. Observe that 
$$
\frac{\partial}{\partial \xi} \, \cF\du(\xi) =
\left[\frac{\partial}{\partial \xi} \cS \right] (\psi_\star)
$$
since the other term, containing 
$\frac{\partial}{\partial \xi} \psi_\star$,  
vanishes by the definition of a maximum. 
Definitions and integration by parts yield 
$$
-\left(
\frac{\partial}{\partial \xi_i} - 
\frac{\partial}{\partial \xi_{i+1}} \right)
\cF\du = \frac{a_i - a_{i+1}}{r} \,. 
$$
By \eqref{Ftrans}, this means that the resulting 
sum over the $r$-quotients $\lambda^{(i)}$
is the original partition function $Z$ with 
parameters $a\in \bA$. This concludes the proof.


\section{The next dimension}\label{sHigh}

\subsection{Stepped surfaces} \quad \newline 

\begin{figure}[!htbp]
  \centering
  \scalebox{0.33}{\includegraphics*[0,185][530,790]{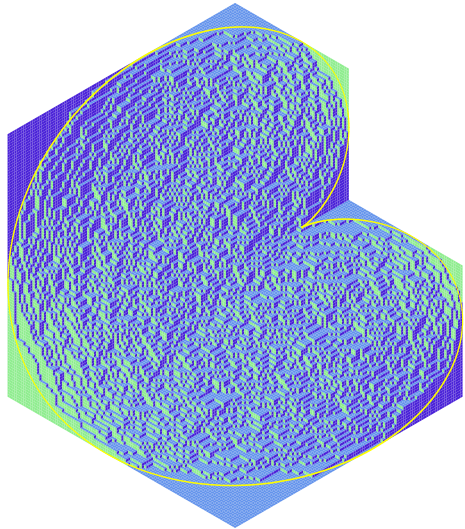}}
  \caption{A limit shape simulation.
The frozen boundary is the inscribed cardioid.}
  \label{fcard}
\end{figure}

\noindent
\begin{minipage}[b]{8.5cm}
An obvious $3$-dimensional generalization of a partition, 
also known as a plane partition can be seen on the right. 
More generally, we consider \emph{stepped surfaces}, that 
is, continuous surfaces glued out of sides of a unit cube, 
spanning a given polygonal contour in $\R^3$,  
and projecting $1$-to-$1$ in the $(1,1,1)$ direction, see
Figure \ref{fcard}.  
Note that stepped surfaces minimize the surface area 
for given boundary conditions, hence
can be viewed as zero temperature limit of the interface
in the 3D Ising model.  
\end{minipage}
\hfill 
\scalebox{0.3}{\includegraphics{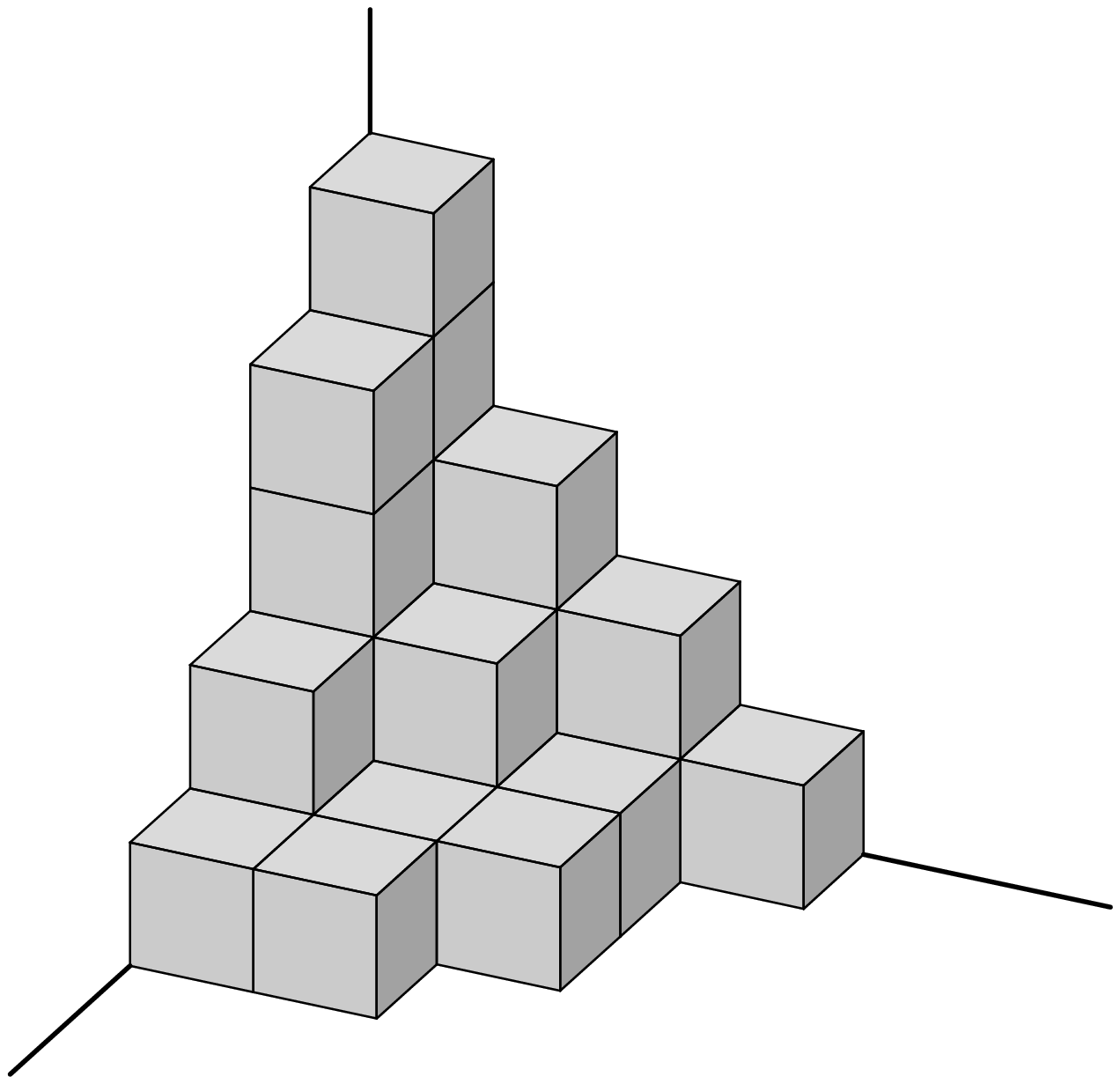}}

The most natural measure on stepped surfaces is the uniform
one with given boundary conditions, possibly conditioned
on the volume enclosed. It induces Plancherel-like 
measures on $2$-dimensional slices. Stepped surfaces 
are in a natural bijection with fully packed dimers on 
the hexagonal lattice and Kasteleyn theory of planar 
dimers \cite{Kas} forms the basis of most subsequent 
developments. 

The following law of large numbers for stepped surfaces
was proven in \cite{CKP}. Let $C_n$ be a sequence of 
boundary contours such that each $C_n$ can be spanned
by at least one stepped surface. Suppose that $n^{-1} C_n$
converge to a given curve $C\subset \R^3$. Then, scaled 
by $n^{-1}$, uniform measures on stepped surfaces spanning 
$C_n$ converge to the $\delta$-measure on a single Lipschitz
surface spanning $C$ --- the limit shape. This limit shape 
formation is clearly visible in Figure \ref{fcard}. 

The limit shape is the unique minimizer of the following 
functional. Let the surface be parameterized
by $x_3=h(x_3-x_1,x_3-x_2)$, where $h$ is a Lipschitz
function with gradient in the triangle $\bigtriangleup$ with vertices
$(0,0),(0,1),(1,0)$. Let $\Omega$ be the planar region 
enclosed by the projection of $C$ in the $(1,1,1)$ 
direction. We will use $(x,y)=(x_3-x_1,x_2-x_1)$ as 
coordinates on $\Omega$. The limit shape is the 
unique minimizer of 
\begin{equation}
  \label{Sstep}
   \cS_\textup{step}(h) = \int_\Omega \sigma_\textup{step}
(\nabla h) \, dx \, dy \,, 
\end{equation}
where, in the language of \cite{KOS},
the surface tension $\sigma_\textup{step}$  
is the Legendre dual of the Ronkin function of the 
straight line
\begin{equation}
  \label{line}
   z +w = 1 \,. 
\end{equation}
We recall that for a plane curve $P(z,w)=0$, its Ronkin function \cite{Mikh}
is defined by
\begin{equation}
  \label{Ronk}
R(x,y) = \frac1{(2\pi i)^2} \iint_{
  \substack{|z|=e^x\\|w|=e^y}} \log \big|P(z,w)\big| \, \frac{dz}{z} \,
\frac{dw}{w}\,.
\end{equation}
The gradient $\nabla R$ always takes values in the Newton polygon
$\Delta(P)$ of the polynomial $P$, so $\Delta(P)$ is naturally
the domain of the Legendre
transform $R^\vee$. For the straight line as above,
the Newton polygon is evidently the triangle $\bigtriangleup$.

The surface tension $\sigma_\textup{step}$ is singular and 
not strictly convex at the boundary of $\bigtriangleup$, which leads
to formation of \emph{facets} and \emph{edges} in the limit shape
(which can be clearly seen in Figure \ref{fcard}). This models 
facet formation in natural interfaces, e.g.\ crystalline 
surfaces, and is the most interesting aspect of the model. 
 Note that facets are completely ordered (or \emph{frozen}). 
The boundary between the ordered and disordered (or \emph{liquid})
regions is known as the \emph{frozen boundary}. 

The following transformation of the Euler-Lagrange equation 
for \eqref{Sstep} found in \cite{Burg} greatly facilitates
the study of the facet formation.   Namely, in the liquid 
region we have 
  \begin{equation}
    \label{arg}
    \nabla h = \frac{1}{\pi} (\arg w, - \arg z)\,, 
\end{equation}
where the functions $z$ and $w$ solve the differential equation
\begin{equation}
  \label{Burg}
  \frac{z_x}{z} + \frac{w_y}{w} = c
\end{equation}
and the algebraic equation \eqref{line}. Here $c$ is the 
Lagrange multiplier for the volume constraint $\int_\Omega h = 
\textup{const}$, the unconstrained case is $c=0$. At the 
boundary of the liquid region, $z$ and $w$ become real and 
the $\nabla h$ starts to point in one of the coordinate 
directions. 

The first-order quasilinear equation \eqref{Burg} is, 
essentially, the complex Burgers equation $z_x = z z_y$ and, 
in particular, it can be solved by complex characteristics
as follows. There exists an analytic function $Q(z,w)$ such 
that 
\begin{equation}
  \label{Q}
    Q(e^{-cx} z, e^{-cy} w)=0 \,.
\end{equation}
In other words, $z(x,y)$ can be found by solving \eqref{line} and 
\eqref{Q}. In spirit, this is very close to Weierstra\ss\ 
parametrization of minimal surfaces in terms of analytic data. 

Frozen boundary can only develop if $Q$ is real, in which 
case the roots $(z,w)$ and $(\bar z,\bar w)$ of \eqref{Q}
coincide at the frozen boundary. At a smooth point of the 
frozen boundary, the multiplicity of this root will be 
exactly two, hence $\nabla h$ has a square-root singularity
there. As a result, the limit shape 
has an $x^{3/2}$ singularity at the generic point of the
frozen boundary, thus recovering the
well-known Pokrovsky-Talapov law \cite{PT} in this
situation. At special points of the frozen boundary, triple
solutions of \eqref{Q} occur, leading to a 
cusp singularity. One
such point can be seen in Figure \ref{fcard}.

Remarkably, for a dense set of boundary condition 
the function $Q$ is, in fact, a polynomial. Consequently, 
the frozen boundary takes the form $R(e^{cx},e^{cy})=0$, 
where $R$ is the polynomial defining the planar dual 
of the curve $Q=0$. This allows 
to use powerful tools of algebraic geometry to study 
the singularities of the solutions, see \cite{Burg}. 
The precise result proven there is 

\begin{theorem}[\cite{Burg}] 
Suppose the boundary contour $C$ is
a connected
polygon with $3k$ sides in coordinate directions
(cyclically repeated) which can be spanned 
by a Lipschitz function with gradient in $\bigtriangleup$. 
Then $Q=0$ is an algebraic curve of degree
$k$ and genus zero.
\end{theorem}

For example, for the boundary contour in Figure 
\ref{fcard} we have $k=3$ (one of the boundary edges 
there has zero length) and hence $R$ is the dual 
of a degree 3 genus 0 curve --- a cardioid. 
The procedure of determining $Q$ from the boundary 
conditions is effective and can be turned into a practical numeric
homotopy procedure, see \cite{Burg}. 
Higher genus frozen boundaries
occur for multiply-connected domains, in fact, 
the genus of $Q$ equals the genus of the liquid
region. 

Of course, for a probabilist, the law of 
large numbers is only the beginning and the 
questions about CLT corrections to the limit
shape and local statistics of the surface in 
various regions of the limit shape follow 
immediately. Conjecturally, the limit shape 
controls the answers to all these questions. 
For example, the function $e^{-cx} z$ defines
a \emph{complex structure} on the liquid 
region and, conjecturally, the Gaussian 
correction to the limit shape is given by the
\emph{massless free field} in the corresponding 
conformal structure. In the absence of 
frozen boundaries and without the volume 
constraint, this is proven in \cite{Ken}. 
See e.g.\ \cite{Jo2,Ken,KOS,randpar,OR} for an introduction to the 
local statistics questions.

\subsection{Periodic weights}\label{sPw}

Having discussed periodically weighted Plancherel 
measure and a $3$-dimensional analog of the Plancherel 
measure, we 
now turn to periodically weighted stepped surfaces. 
This is very natural if stepped surfaces are 
interpreted as crystalline interfaces. Periodic
weights are introduced as follows: we weight each 
square by a periodic function of $x_3-x_1$ and 
$x_2-x_1$ (with some integer period $M$). 

The role previously played by the straight line 
\eqref{line} is now played by a certain higher 
degree curve $P(z,w)=0$, the spectral 
curve of the corresponding periodic Kasteleyn 
operator. In particular, the surface
tension $\sigma_\textup{step}$ is now 
replaced by the Legendre dual of the 
Ronkin function of $P$, see \cite{KOS}.
We have
$$
\deg P = M 
$$ 
and the coefficients of $P$ depend 
polynomially on the weights.

The main result of \cite{KOS}, known as
\emph{maximality}, says that for 
real and positive weights the curve $P$ is 
always a real algebraic curve of a very 
special kind, namely, a \emph{Harnack curve},
see \cite{Mikh}. Conversely, as shown in 
\cite{KO1}, all Harnack curves arise in 
this way. 
\begin{figure}[!hbtp]
  \centering
\scalebox{0.35}{\includegraphics{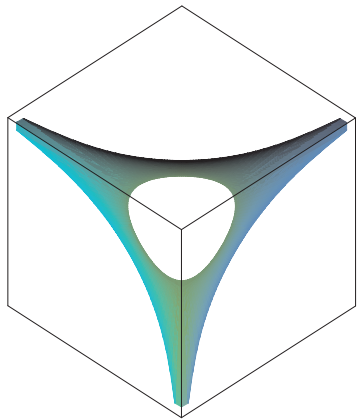}}
\caption{The (curved part of the) Ronkin function of a genus $1$
curve. Its projection to the plane is the amoeba.}
  \label{fRonk}
\end{figure}

Harnack curves are, in some
sense, the best possible real curves; their 
many remarkable properties are discussed in 
\cite{Mikh}. One of several equivalent 
definitions of a Harnack curve is that the 
map 
\begin{equation}
  \label{amoeb}
   (z,w) \mapsto (\log |z|, \log |w|)
\end{equation}
from $P(z,w)=0$ to $\R^2$ is $1$-to-$1$ on 
the real locus of $P$ and $2$-to-$1$ over the rest. 
The image of $P=0$ under \eqref{amoeb} is 
known as the \emph{amoeba} of $P$. Note from 
\eqref{Ronk} that the gradient $\nabla R$ of 
the Ronkin function of $P$ is nonconstant precisely
for $(x,y)$ in the amoeba of $P$. In other words, 
the Ronkin function has a facet (that is, a linear
piece) over every component of the amoeba 
complement. The $2$-to-$1$ property implies that 
the number of compact facets of Ronkin function 
equals the (geometric) genus of the curve $P$.
Each of these facets translates into the singularity 
of the surface tension and, hence, into facets with 
the same slope in limit shapes. 

By Wulff's theorem, the Ronkin function itself 
is a minimizer, corresponding to its own (``crystal
corner'') boundary conditions. An example of 
the Ronkin function of a genus $1$ Harnack curve 
can be seen in Figure \ref{fRonk}. 

Maximality implies \emph{persistence of facets}, 
namely, for fixed period $M$, there will be 
$\binom{M-1}{2}$ compact facets of the Ronkin 
function and $\binom{M-1}{2}$ corresponding singularities of the surface tension,
except on a 
codimension $2$ subvariety of the space of weights. 
It also implies e.g.\ the following 
\emph{universality of height fluctuations} in the 
liquid region
$$
\Var(h(a)-h(b)) \sim \frac{1}{\pi} 
\ln \, \|a-b\| \,, \quad \|a-b\|\to\infty\,. 
$$

Remarkably, formulas \eqref{arg}, \eqref{Burg}, and 
\eqref{Q} need no modifications for periodic weights. 
Replacing \eqref{line} by $P(z,w)=0$ is the only 
change required, see \cite{Burg}. 

{}From our experience with periodically weighted 
Plancherel measure, it is natural to expect that,
for some special 
boundary conditions, 
the partition function of periodically weighted 
stepped surfaces will encode valuable 
physical information. A natural choice of 
``special boundary conditions'' are the those of 
a \emph{crystal corner}, when we require the 
surface to be asymptotic to given planes at 
infinity, as in Figure \ref{fRonk}. For 
convergence of the partition function, one 
introduces a fugacity 
factor $q^{\textup{vol}}$, where the missing 
volume is measured with respect to the 
``full corner''. 

\begin{figure}[!hbtp]
  \centering
\rotatebox{90}{\scalebox{0.4}{\includegraphics{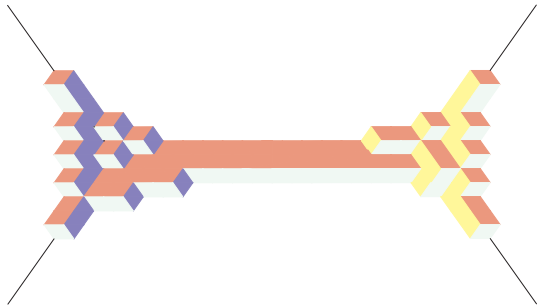}}}
\caption{Two 3D partitions connected at an angle through
an infinite leg.}
  \label{fedge}
\end{figure}

I hope that further study will reveal many
special properties of such crystal 
corner partition functions. Their extremely
degenerate limits 
have been identified with 
all-genera, all-degree generating
functions for Donaldson-Thomas invariants
of toric Calabi-Yau threefolds. Namely, 
as the periodic weights become extreme, 
all limit shapes, and the Ronkin function in 
particular, degenerate to piecewise linear
functions. This is known as the \emph{tropical
limit}. The only remaining features of limit shapes  
are the edges 
and the triple points, where $2$ and $3$ 
facets meet, respectively. In this 
tropical limit, the partition function 
becomes the partition function of ordinary, 
unweighted, 3D partitions located at 
triple points. These 3D partitions may 
have infinite legs along the edges, as 
in Figure \ref{fedge} and through these
legs they interact with their neighbors. 
This description precisely matches the 
localization formula for Donaldson-Thomas
invariants of the toric threefold whose
toric polyhedron is given by the 
piecewise linear limit shape, see \cite{mnop}.

Donaldson-Thomas theory of any $3$-fold
has been conjectured to be equivalent, 
in a nontrivial way, to the Gromov-Witten 
theory of the same $3$-fold in \cite{mnop}. 
For the toric Calabi-Yau $3$-folds, this 
specializes to the earlier \emph{topological 
vertex} conjecture of \cite{vafa}. It is impossible 
to adequately review this subject here, see \cite{EC} for 
an introduction. This is also related to 
the supersymmetric gauge theories considered
in Section \ref{sGau}, or rather their 5-dimensional
generalizations, via a procedure 
called \emph{geometric engineering} of gauge
theories. See for example \cite{strip} and 
references therein. 

I find such close and unexpected 
interaction between rather basic 
statistical models and instantons in 
supersymmetric gauge and string theories
very exciting and promising. The field 
is still full of wide open questions 
and, in my opinion, it is also 
full of new phenomena waiting to be 
discovered.

\end{document}